\documentclass{rmaa}
\def\spose#1{\hbox to 0pt{#1\hss}}
\def\lta{\mathrel{\spose{\lower 3pt\hbox{$\mathchar"218$}}
     \raise 2.0pt\hbox{$\mathchar"13C$}}}
\def\gta{\mathrel{\spose{\lower 3pt\hbox{$\mathchar"218$}}
     \raise 2.0pt\hbox{$\mathchar"13E$}}}
\usepackage{rmaacite}
\def\bmath#1{\mathbf{#1}}

\input epsf  

\title{N-body simulations of small galaxy groups}
\author{H\'ector Aceves\altaffilmark{1} and H\'ector Vel\'azquez
\affil{Instituto de Astronom\'{\i}a UNAM. Ensenada, M\'exico. }
}
\altaffiltext{1}{E-mail: \texttt{aceves@astrosen.unam.mx}}
\fulladdresses{
 \item H\'ector Aceves: Instituto de Astronom\'{\i}a, UNAM. Apdo. Postal 877.  Ensenada B.C. 22860. M\'exico.} 
\shortauthor{Aceves  \& Vel\'azquez}
\shorttitle{Groups of galaxies}

\resumen{
Se presenta una serie de simulaciones de $N$-cuerpos destinadas a 
estudiar la din\'amica de grupos peque\~nos de galaxias. En particular, 
se comparan los resultados obtenidos con las propiedades din\'amicas de los 
grupos de Hickson. 
Se consideran condiciones iniciales partiendo de una `expansi\'on m\'axima' y 
de equilibrio virial, 
y no se considera un halo oscuro com\'un primigenio.
   Se encuentra muy buena concordancia con las propiedades de grupos peque\~nos
de galaxias, y  las propiedades de los grupos compactos de Hickson son reproducidas adecuadamente por aquellos sistemas que se encuentran en un estado avanzado de un colapso gravitacional.
  No se encuentra un problema de `sobre-fusi\'on' de las 
galaxias. Para grupos que comienzan en virial, se encuentra que una fracci\'on importante ($\sim 40$\%) pueden 
durar por $\sim 10$ Gyr sin colapsarse completamente. Los resultados encontrados proporcionan otra 
soluci\'on alternativa al problema  de sobre-fusi\'on para los grupos de Hickson. 
Asimismo, se encuentra que la raz\'on masa-luminosidad de los grupos de Hickson podr\'{\i}a ser similar a la de 
c\'umulos de galaxias, sugiriendo  que ambos tienen aproximadamente la misma 
fracci\'on de materia bari\'onica a masa total.
}
\abstract{
A series of $N$-body simulations aimed to study  the dynamics of small groups of galaxies are presented. 
In particular, our results are compared with the dynamical properties of Hickson's compact groups (HCG).   
 `Maximum expansion' and virial initial 
conditions are tested, and no primordial common dark halo is considered. The properties of small galaxy groups are very well reproduced, and those of Hickson's groups are well reproduced by the most advanced stage of collapsing  groups. 
 We find no overmerging problem in our simulations. An important fraction of groups ($\sim 40$\%) initially in virial equilibrium can last for 
$\sim 10$ Gyr without complete merging. These results  provide  an alternative solution to the 
overmerging expected in Hickson's compact groups. Also, the mass-to-light ratio of 
HCG are probably similar to those found in clusters, suggesting that both kinds of systems have about 
the same fraction of barionic to total mass.
}

\begin{document}
\maketitle

\section{Introduction}

Small groups are the most common galaxy associations and contain about
$50$\% of all galaxies in the universe \cite{huchra-geller,nolthenius}.  
Important work has been done to compile catalogs with relevant kinematical 
data. Thus Nolthenius \& White (1987) and Nolthenius 
(1993) found that small groups from the CfA catalog (hereafter NCfA) 
have the following median values: a one-dimensional velocity dispersion of 
$\sigma \approx 116$  
km s$^{-1}$, a mean harmonic radius of $R_{\rm H}= 480$ kpc, a deprojected median 
radius of $R_{\rm S}=720$ kpc and a crossing time of $H_0\tau_{\rm c} = 0.44$.\footnote{Here, $R_{\rm S}\equiv 4 R/\pi$,   $R_{\rm H}^{-1}\equiv 
4 \Sigma_i \Sigma_{i<j}\,R_{ij}^{-1}/(\pi N_{\rm g}(N_{\rm g}-1))$,
and $\tau_{\rm c}\equiv  2R_{\rm H}/(\sqrt{3}\sigma)$,
 where $R$, $R_{ij}$ 
and $N_{\rm g}$ are the average projected separation, the projected separation 
between galaxy pairs and the number of galaxies, respectively.} Hereafter 
a Hubble constant of $H_0=75$ km s$^{-1}$Mpc$^{-1}$ is assumed. 
Gourgoulhon, Chamaraux 
\& Fouqu\'e (1992, hereafter GCF) found for nearby small 
groups ($<80$ Mpc) a median 
$\sigma\approx 75$ km s$^{-1}$, a virial radius of about $770$ kpc and a 
crossing time-scale of $\sim 4$ Gyr,  these quantities corrected for 
the effect of the Hubble flow. More recently, Makarov \& Karachentsev (2000, hereafter MK) 
derived from a sample of 839 groups, in particular for those consisting of five galaxies, the 
following mean values, with quartiles: $\sigma = 57^{+26}_{-14}$ km s$^{-1}$, a non-deprojected 
harmonic radius of $230^{+92}_{-99}$ kpc and a dimensionless crossing time of about $0.1$. 
   
A special and very important case is Hickson's compact groups (HCG)
which are characterized for being composed of at least 
four galaxies with a median projected separation of about three times the 
diameter of their luminous component and densities similar to the central 
region of rich clusters \cite{hickson82,hickson97}. 
Hickson's initial catalog contained 
100 groups with a median projected separation of $R=52_{2.1}^{180}$ kpc, a velocity dispersion of $\sigma =200_{13}^{617}$ km s$^{-1}$ and a 
dimensionless crossing time of $H_0 t_{\rm c}=0.016_{0.001}^{8.7}$; 
where $t_{\rm c}
\equiv 4R/(\pi \sqrt{3}\sigma)$, and subscripts and superscripts indicate the range of observed values. However, Sulentic (1997) pointed out 
that only $61$ of them show according redshifts. 

It is noteworthy that HCG provide an ideal site to study 
the evolution of galaxies, via 
interactions and/or mergers, in a high density environment.  
One of the most striking challenges of the compact groups is how to reconcile 
their apparent longevity  with their quite small crossing time of 
about $1$\% of the Hubble time. This short time-scale suggests a 
rapid evolution of these systems toward a complete merging: the so-called 
{\it overmerging} problem (Hickson, Richstone \& Turner 1977; Hickson 1982; White 1990; Hickson 1997). However, Hickson et al.~(1984) and Rubin, Hunter \& Ford (1991) have suggested that compact groups may have formed relatively recently. 

In general, two complementary approaches have been proposed to address this 
issue; see Athanassoula (2000) for review on this topic. On the one hand, 
compact groups are assumed to be formed in the early universe and survive 
somehow until the present epoch (Ishizawa 1986; Diaferio, Geller \& 
Ramella 1994; Governato, Tozzi \& Cavaliere 1996). Diaferio et al. (1994) 
introduced a scenario where a {\it continuous} replenishment of the 
compact group occurs within a {\it single} collapsing rich loose group. 
Governato et al. (1996) proposed an alternative model that relies in 
the evolution of a compact group as an ongoing and frequent process 
through  secondary infall of galaxies in a critical universe. 
However, in these scenarios merging is not suppressed so one would expect 
more compact groups and merger remnants than those observed today 
\cite{mamon2000}. 

On the other hand, there is a dynamical view where some kind of tuning 
of the initial conditions is imposed in order to guarantee the survival of 
the compact group. In this sense, Barnes (1985) was the first to 
investigate the effects of different kinds of initial conditions 
on the evolution of compact groups. He found that (1) a compact group 
immersed in a {\it massive common halo} would, in general, delay its merging 
process if more mass is initially placed in this common halo. This 
merging delay results from the fact that the halo common mass 
is obtained from the galaxy haloes, thus reducing their masses 
and increasing the dynamical friction time-scale. This result was also 
corroborated by Bode, Cohen and Lugger (1993). 
And, (2), groups whose initial conditions begin in `turnaround' or 
`expansion' merge more rapidly than their virialized counterparts.

More recently, Athanassoula et al.~(1997) carried out a more 
systematic study of the parameter space for small groups 
(e.g.~concentrations of the common and individual haloes, common-halo-mass 
 to total-group-mass ratios, etc.).~Their main findings are: (1) 
centrally concentrated groups merge faster for high common-halo to 
total-mass ratios and slower for low common-halo to total-mass ratios. (2) 
The overmerging issue of compact groups may not be a problem if 
{\it appropriate} initial conditions are chosen. To support this last 
point, they built up a virialized group with a large common-halo to 
total-group and an almost homogeneous central concentration that was able to 
survive for as long as $\sim \, 20$ Gyr. According to these results the merger 
rates obtained from group simulations without including a common envelope, as the ones to be considered here, would be an upper limit.

In this work the hypothesis of Barnes (1989) and White (1990) regarding that diffuse groups might be the progenitors of compact ones, but criticized by Diaferio~et~al.~(1994) for lacking quantitative support, is tested; similarly, the hypothesis  advanced on observational grounds that HCG might be relatively young systems (Hickson~et~al.~1984, Rubin~et~al.~1991).
The dynamical approach is used, and for this end a series of N-body 
simulations are performed {\it without} including a common halo. 
These simulations allow us to 
compare with the observed kinematical properties of small groups. The rest 
of this paper has been organized as follows: in section $\S 2$  
the numerical methods to set up the galaxy models and the initial 
condition for the group are described.
 In section $\S 3$ the merging 
histories of our simulations and the dynamical properties are given in 
section $\S 4$. In section $\S 5$ the mass estimation of our 
groups is addressed. A constrained representation of the `phase-space' is 
shown in section $\S 6$. A more realistic N-body simulation involving 
spiral galaxies is given in section $\S 7$, and section $\S 8$ contains a 
general discussion of the results. Finally, the main conclusions are 
summarized in section $\S 9$.

\section{Description of the Simulations}

This section contains a detailed description about the method used to 
set up our galaxy model and the initial conditions for our small galaxy 
groups. Also, the computational tools and the parameters used 
to evolve our simulations are described.

\subsection{Galaxy Model}

Individual galaxies are represented by a self-consistent Plummer model.  
In building up the galaxy model no explicit difference 
between dark and luminous matter is done. 
However, to gain some feeling about the 
evolution of a luminous component, the initial $10$\% of most 
bounded particles are identified as `luminous'.

By using the procedure described by Aarseth, H\'enon and Wielen (1974), 
positions and velocities of the particles are randomly generated from 
the following mass and phase-space distributions $f({\cal E})$:

$$
M(r) = M_{\rm g}\frac{(r/R_0)^3}{[1+ (r/R_0)^2]^{3/2}} \;
$$
and
$$
f({\cal E}) = \frac{24 \sqrt{2} R_0^2}{7 \pi^3 G^5 M_{\rm g}^4} |{\cal{E}}|^{7/2},
$$
where $M_{\rm g}$ is the total galaxy mass, $R_0$ its scale-length, 
and $\cal{E}$ the energy per unit mass, respectively.

A system of model units such that $G=1$,~$M_{\rm g}=1$ and $R_0=1$ is chosen.~Since at large radii the Plummer density goes as $\rho \propto r^{-5}$  a cut-off radius is introduced, for numerical purposes, at about 
$r_{\rm{cut}}\approx 10\,R_0$ which contains about  $99$\% of the total galaxy 
mass. In this system of units, the 
dynamical time-scale, the half-mass and the cut-off radii are given by 
$t_{\rm d}=\sqrt{8R_0^3/(GM_{\rm g})}=\sqrt{8}$, $R_{\rm h}\approx 1.3$ and 
$r_{\rm{cut}}=10$, respectively. In order to compare with observational data the values  
$r_{\rm{cut}}\approx 135$ kpc and $M_{\rm g}\approx 5.5 \times 10^{11}$ 
M$_\odot$ (Model B of Kuijken \& Dubinski (1995)) are adopted.
 For these values the 
time and velocity units are $31.5$ Myr and $419$ km s$^{-1}$. 
In general, the simulations can be easily scaled through the following 
expressions for the velocity and time: 
$v=2.1 \times 10^{-3}\sqrt{M_{\rm g}/R_0}$ km s$^{-1}$ and $t=4.7\times10^{5}
\sqrt{R_0^3/M_{\rm g}}$ Myr, where $R_0$ and $M_{\rm g}$ must be given in kpc 
and M$_\odot$, respectively. With these definitions our initial 
`luminous' core is contained within a radius of $0.74$ model units,  
$\approx 10$ kpc. Finally, each galaxy is represented by $3000$ 
equal-mass particles.

\subsection{Group Initial Conditions}

In all cases each group consists of five equal-mass galaxies. 
Galaxies are assumed to be already formed and no secondary infall 
is considered, which is justifiable 
in a low density universe (e.g. Bahcall 1999, Hradecky 
et al.~2000). The positions and velocities for the center of mass 
of these galaxies are randomly generated from initial conditions starting 
at `turnaround' and at virial equilibrium. In all simulations 
no primordial common dark halo is included. Although this 
assumption seems to contradict current cosmological models of structure 
formation, some observations on clusters and groups suggest that most of the 
dark matter is associated with the individual dark haloes of galaxies 
(e.g. Bahcall, Lubin \& Droman 1995, Puche \& Carignan 1991). Hence, 
 the merging activity observed in our simulations could be 
an upper limit.

\subsubsection{Collapsing groups}

For groups initially in `turnaround' the positions for the 
centers of galaxies were obtained randomly from an homogeneous  sphere 
of radius $R_{\rm max}$ defined by 
(Gunn \& Gott 1972, ignoring any contribution from the cosmological constant)
$
R_{\rm {max}} = (8 G M_{\rm G} t^2/\pi^2)^{1/3} ,
$
where $M_{\rm G}$ represents the mass of the group enclosed within this radius 
at time $t$; mass conservation will be assumed. 
 {\it  Compact groups are assumed to be at the verge of complete collapse 
at the present epoch} (e.g. Hickson et al.~1984, Rubin et al.~1991), 
then their collapsing time-scale $t_{\rm clps}$ turns to be the age 
of the universe $t_0$. Hence, $R_{\rm max}$ was evaluated at $t_0/2$ (Gott \& Rees 1975).

For 
$\Omega_0=1$ it is found that $t_0=2H_0^{-1}/3$, and for $\Omega_0 \to 0$ that
 $t_0\approx
H_0^{-1}(1+0.5\Omega_0\ln\Omega_0)$ (Gott et al. 1974).\footnote{This approximation differs by less than $1$\% from its true value 
for $\Omega_0=0.2$ (Peacock 1999).} In particular for $\Omega_0=0.2$ and $H_0=75$ 
km s$^{-1}$Mpc$^{-1}$  $t_0\approx 10$~Gyr the `turnaround' epoch was 
$5$~Gyr ago (i.e.~at redshift $z\approx0.6$, Sandage 1961). 
In Table 1  the 
`turnaround' radius of the group is listed for different values of $H_0$ and 
$\Omega_0$. A $R_{\rm max}=700$ kpc is taken as the fiducial `turnaround' radius for collapsing groups. Notice that this value is consistent with a 
$H_0=50$ km s$^{-1}$Mpc$^{-1}$ and $\Omega_0=1$, and with the observational 
data of Sulentic (1987) who found that there is not a significant 
population of normal galaxies within $\sim 1$~Mpc of the central region of 
HCG.

Finally, the bulk velocity magnitudes for the galaxy centers were obtained randomly from a 
Gaussian distribution with a one-dimensional velocity dispersion of 
$\sigma=\sqrt{4GM_G/(5R_{\rm max})}$ (Gott 1975) subject to the 
constrain $2T/|W|=1/4$ (Barnes 1985), where $T$ and $W$ are the kinetic and potential energy of the group, respectively; these energies were computed assuming that galaxies are equal point-mass particles. Velocity vectors were oriented randomly.

\subsubsection{Virialized groups}

Theoretical arguments and $N$-body simulations (Peacock 1999) indicate that
by a time $\approx 3 t_{\rm clps}/2$ a system has reached an approximate equilibrium state with a virial radial scale of $R_{\rm v}\approx R_{\rm max}/2$.
Hence a radial scale for groups in near virial equilibrium at $t_0$ is
$R_{\rm v}=R(3t_0/2)/2$, and in the particular case for 
our fiducial model a $R_{\rm v}=267$~kpc is obtained. The positions for the 
galaxies are randomly chosen from a uniform spherical distribution with 
radius $R_{\rm v}$, and their velocities from a Gaussian distribution  restricted to satisfy the virial ratio $2T/|W|=1$ with random orientation vectors.

\subsection{Computational Issues}

To evolve our galaxy and group models a serial {\rm TREECODE} was used 
(Barnes \& Hut 1986) with a tolerance parameter $\theta=0.9$ and with the 
quadrupole corrections to the potential included. A softening 
parameter $\epsilon=0.07 \approx R_0/14$  and a time-step $\Delta t=0.1$ were
adopted. With 
these parameters, the simulations  were evolved for about $321.4$ time units 
($\sim10$ Gyr) and the energy conservation, in all cases, was better than 
$0.7$\%. A total set of $30$ simulations were performed for each kind of initial conditions considered for the groups.

To follow up the orbital evolution of each galaxy inside the group  
 its center was identified with its most bounded particle, which was set by 
`hand' to have $1$\% of the galaxy mass (Aguilar \& White 1985). In order to check the stability 
of the galaxy model due to this change its evolution was followed, in 
isolation, for $\sim 200$ dynamical time-scales,  no 
 significant change in its inner structure and in the total energy 
conservation was found. From here on, all kinematical properties of the group 
simulations  (e.g. mean separation, 
velocity dispersion, etc.) will be obtained from these galaxy centers.

\begin{table}
\caption[]{Turnaround radius (Mpc)}
\label{tab:rmax}
\begin{center}
{\small
\vspace{-0.3cm}
\begin{tabular}{ccc}
\hline\noalign {\smallskip}
 $H_0$    & $\Omega_0=0.2$  & $\Omega_0=1$    \\
(km s$^{-1}$Mpc$^{-1}$) &  & \\
\noalign{\smallskip}
\hline
\noalign{\smallskip}
50  &  0.88 & 0.76  \\
75  &  0.67 & 0.58  \\
100 &  0.56 & 0.48  \\
\noalign{\smallskip}
\hline
\end{tabular}    
}
\end{center}
\end{table}

\begin{figure*}
\begin{center}
 \epsfxsize=16cm
\end{center}
\caption{This figure shows the XY-projection in kpc for the collapsing 
groups $g27c$ ({\it upper-left}), $g02c$ ({\it upper-right}) and 
$g22c$ ({\it bottom-left}). The numbers in each frame indicate the time 
elapsed since turnaround in Gyr. Lines refer to the trajectories followed by 
each galaxy inside its corresponding group. Also,  small boxes of 300 kpc 
wide  are shown at time $4.79$ Gyr for each group. These boxes have been 
amplified  and show just 'luminous' particles at the {\it bottom-right} 
of this figure which corresponds to the XY, XZ and YZ-projections. 
}
\label{fig:3groups_proj}

\end{figure*}

\section{Merging Histories}

In this section the merging histories for all our simulations of 
small groups are computed. These histories provide some insight about the merging rate 
and the overmerging problem of compact groups posed by their small crossing 
time. 

In Figure~\ref{fig:3groups_proj} the evolution of three particular simulations with cold initial conditions is shown.  Notice that 
group $g27c$ does not present any merger during the simulation ($\sim 10$~Gyr) 
while group $g02c$ completely merged. At present epoch (i.e.~$\sim 5$~Gyr 
from `turnaround'), group $g22c$ resembles a chain-like compact group which 
is better appreciated at the {\it bottom-right} of this figure where only 
`luminous' particles have been displayed. 
Notice also that groups $g27c$ and $g02c$ may be classified as triplets.

In order to characterize this merging activity  the following 
pair-wise merging criteria (Aarseth \& Fall 1980) are employed:
$$
V_{ij}  = |\bmath{v}_j - \bmath{v}_i |  <   V_{\rm rms}/2 , \;\,
R_{ij} = |\bmath{r}_j - \bmath{r}_i | < R_0 /2 \,;
$$
where $V_{ij}$ and $R_{ij}$ are the relative velocity and separation between 
a pair of galaxies $i$ and $j$, respectively, and $V_{\rm rms}$ and $R_0$ are 
their characteristic internal velocity and radius.

\begin{figure}
\epsfxsize=8.3cm
\begin{center}
\epsffile{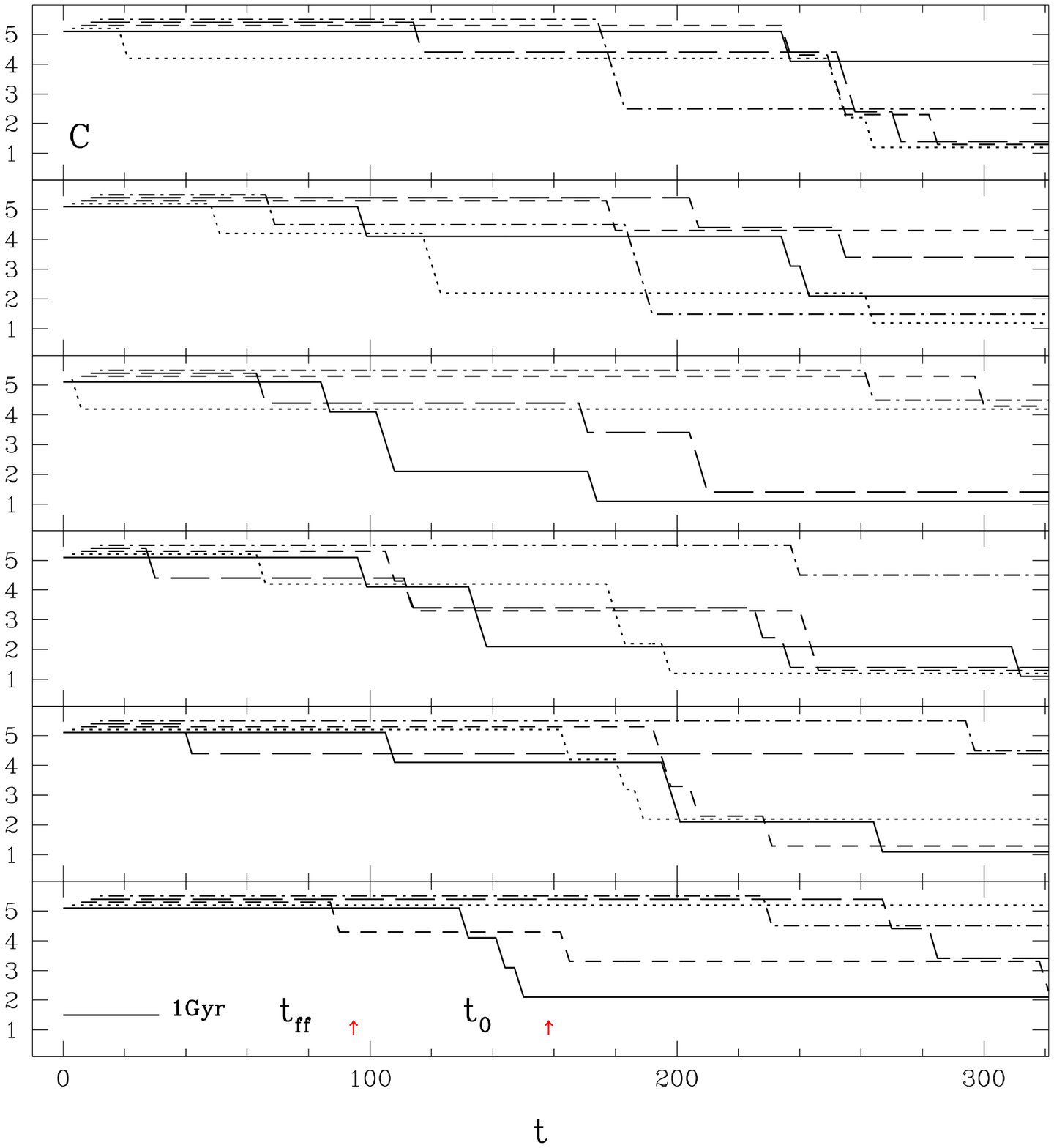}
\end{center}
\vspace{-0.7cm}
\caption{ Number of individual galaxies versus time (merging history)
 for simulations  with cold initial conditions. 
The free-fall time-scale \cite{mamon90} $t_{\rm ff}=\pi\sqrt{R_{\rm rmax}^3/(32GM_{\rm G})}$  and the present epoch $t_0$ ($ \sim 5 $ Gyr from `turnaround') 
are indicated by arrows. The merging tracks have 
been displaced vertically and in time for better appreciation. The 
displacement in time corresponds to 0, 3, 6, 9 and 12 time units. A 
scale corresponding to $1$ Gyr is indicated.
}
\label{fig:merging-c}
\end{figure}
\begin{figure}
\epsfxsize=8.3cm
\begin{center}
\epsffile{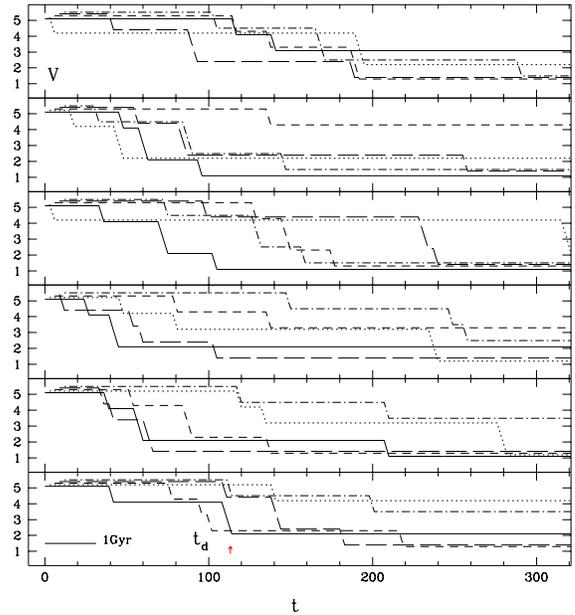}
\end{center}
\vspace{-0.7cm}
\caption{ As in Fig.~\ref{fig:merging-c}, but for initially virialised groups. 
In this case, the arrow corresponds to the dynamical time of the group.}
\label{fig:merging-v}
\end{figure}

In Figures~\ref{fig:merging-c} and~\ref{fig:merging-v}  the time 
evolution of the merging activity in our  groups is shown.  The dynamical 
time for the virialised groups is 
$t_{\rm d}=\sqrt{8R_{\rm v}^3/(GM_{\rm G})}\approx 111$ corresponding 
to about 3.6 Gyr. 
It is observed in Fig.~\ref{fig:merging-c} that 
{\it no} group has merged 
completely at the present epoch, $t_0$, although some multiple mergers do
 exist. 
  About $15$  groups ($50$\%) do not present any merger at all, 
$9$ ($30$\%) show a merger of two galaxies and $6$ ($20$\%)  triple or quadruple mergers. Notice 
that at the end of the simulations just one group does not show any merger and 
 15 ($50$\%) have merged completely.

In the case of virialized groups, 5 ($\approx 17$\%) do not present any kind of merging 
within a dynamical time-scale, 11 ($\approx 37$\%) show a  merger of two galaxies, 10 
 ($\approx 33$\%) have triple 
or quadruple mergers, and only 4 ($\approx 13$\%) have completely merged (see figure~\ref{fig:merging-v}). Notice that about 12 ($40$\%) 
 have {\it not} completely merged after $\sim 10$ Gyr of evolution. 
This clearly indicates that some compact groups can survive complete merging 
for about a Hubble time.

\section{Dynamical Properties}

The dynamical properties of the groups 
along three orthogonal `lines-of-sight' at different times are determined. The median values for 
the one-dimensional velocity dispersion, the mean harmonic and deprojected 
radii, the VME and MME masses (see $\S 5$), and the dimensionless crossing 
time, $H_0 \tau_{\rm c}$, and the dimensionless crossing time as in Hickson et al.~(1992), $t_{\rm c}$,
are summarized in Table~\ref{tab:res-tab2}. The times indicated in 
the table correspond to $0.5$, $1$, $2$, ..., $10$ Gyr of evolution from their 
respective initial conditions.

\begin{table*}
\caption{Dynamical properties of groups in model units}  
\label{tab:res-tab2}
\begin{center}
\vspace{-0.5cm}
{\small   
\begin{tabular}{r|crrcrll|crrcrll}
\hline 
      & \multicolumn{7}{c|}{Collapsing} & \multicolumn{7}{c}{Virialized} \\
t     & $\sigma$ & $R_{\rm H}$ & $H_0\tau_{\rm c}$ & $R_{\rm S}$ & $H_0 t_{\rm c}$ & $M_{\rm v}$ & $M_{\rm med}$  & $\sigma$ & $R_{\rm H}$ & $H_0\tau_{\rm c}$ & $R_{\rm S}$ & $H_0 t_{\rm c}$ & $M_{\rm v}$ & $M_{\rm med}$ \\
\hline
15.0  & 0.071 & 45.76 & 1.91 & 51.27 & 1.09 & 1.28 & 0.65   & 0.221 & 15.14 & 0.197 & 19.17 & 0.126 & 3.41 & 1.79  \\
36.0  & 0.077 & 43.25 & 1.57 & 49.74 & 0.99 & 1.50 & 0.67  & 0.223 & 10.79 & 0.126 & 17.18 & 0.111 & 2.71 & 2.06\\ 
63.0  & 0.090 & 36.09 & 1.01 & 46.78 & 0.72 & 1.46 & 0.79  & 0.189 &  2.72 & 0.043 & 18.65 & 0.134 & 0.99 & 1.49 \\ 
96.0  & 0.112 & 25.65 & 0.43 & 41.19 & 0.49 & 1.33 & 0.62 & 0.175 &  1.44 & 0.033 & 20.10 & 0.179 & 0.27 & 0.66  \\ 
126.0 & 0.132 &  8.41 & 0.11 & 34.04 & 0.33 & 0.88 & 0.92  & 0.141 &  0.55 & 0.015 & 21.23 & 0.231 & 0.06 & 1.22  \\ 
156.0 & 0.149 &  3.70 & 0.06 & 27.86 & 0.28 & 0.76 & 1.13  & 0.137 &  0.45 & 0.012 & 20.31 & 0.212 & 0.06 & 0.60  \\ 
186.0 & 0.153 &  0.87 & 0.02 & 22.31 & 0.21 & 0.16 & 1.74  & 0.111 &  0.24 & 0.008 & 21.47 & 0.195 & 0.02 & 0.08 \\ 
219.0 & 0.158 &  0.53 & 0.02 & 17.21 & 0.19 & 0.11 & 1.10  & 0.111 &  0.21 & 0.008 & 21.10 & 0.241 & 0.01 & 0.04  \\ 
252.0 & 0.152 &  0.28 & 0.01 & 15.07 & 0.17 & 0.04 & 0.95  & 0.082 &  0.15 & 0.007 & 20.66 & 0.239 & 0.01 & 0.02\\ 
282.0 & 0.131 &  0.23 & 0.01 & 15.59 & 0.20 & 0.03 & 0.57  & 0.077 &  0.15 & 0.007 & 18.39 & 0.242 & 0.01 & 0.01\\ 
315.0 & 0.100 &  0.19 & 0.01 & 17.20 & 0.21 & 0.02 & 0.14 & 0.083 &  0.11 & 0.005 & 15.41 & 0.197 & 0.01 & 0.00 \\ 
\hline
\end{tabular}
}
\end{center}
{\small
{\scriptsize $t$} time evolved from the initial conditions. 

{\scriptsize $\sigma$} one-dimensional velocity dispersion.

{\scriptsize $R_{\rm H}$} harmonic radius.

{\scriptsize $R_{\rm S}$} mean separation radius.

{\scriptsize $M_{\rm v}$} virial mass.

{\scriptsize $M_{\rm med}$} median mass.

{\scriptsize $H_0t_c$} crossing-time scale where {\scriptsize $t_{\rm c}\equiv 
R_{\rm S}/(\sqrt{3}\sigma)$}.

{\scriptsize $H_0\tau_c$} crossing-time scale where 
{\scriptsize $\tau_{\rm c}\equiv 2R_{\rm H}/(\sqrt{3}\sigma)$}.
}
\end{table*}

\subsection{Collapsing Groups}

\begin{figure*}[!t]
\epsfxsize=17cm
\begin{center}
\epsffile{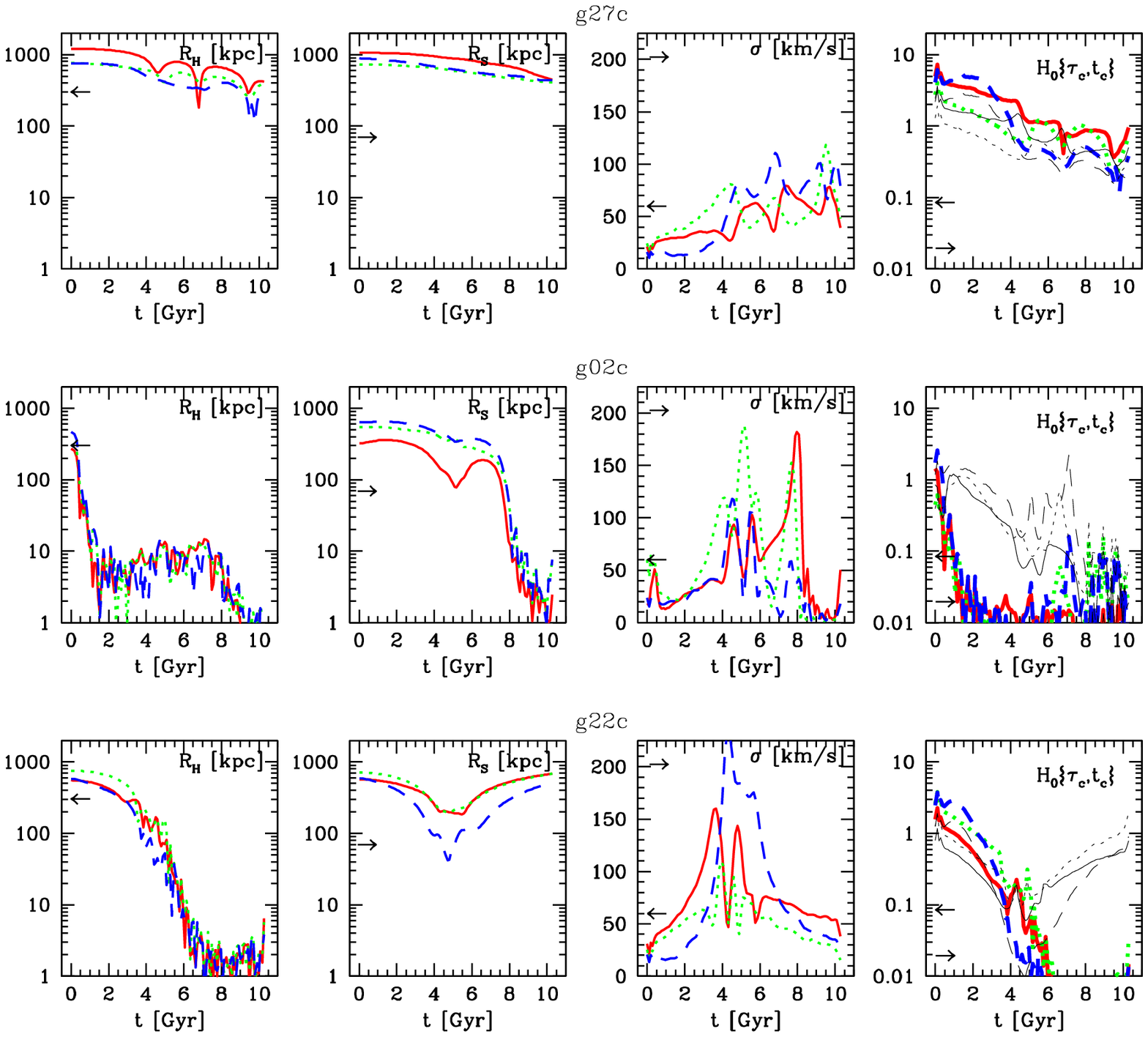}
\end{center}
\vspace{-2.5cm}
\caption{Kinematical parameters as a function of time for collapsing groups $g27$, $g02c$, and $g22c$. Astronomical units are used. The left-arrow ($\leftarrow$) indicates the median of the 5-galaxy MK groups, while the right-arrow ($\rightarrow$), HCG. The different lines indicate the line-of-sight used to compute the quantities: along X-axis (solid), along Y-axis (dotted), and along Z-axis (broken). For the dimensionless crossing times, thicker lines for $H_0 \tau_{\rm c}$ than for $H_0 t_{\rm c}$  have been used.}
\label{fig:3groups_kin}
\end{figure*}

In Figure~\ref{fig:3groups_kin} the dynamical parameters as a function 
of time for our collapsing groups $g27c$, $g02c$ and $g22c$ are shown, 
respectively. 
For comparison purposes the medians of 
MK and HCG are indicated by arrows. This figure shows the general trend in a collapse; a decrease in 
size followed by an increase in velocity dispersion and a reduction in the 
crossing time. The size of a group depends on the method adopted; the harmonic 
radius is always lower than the mean separation radius and, hence, 
$\tau_{\rm c} < t_{\rm c}$. An agreement between both definitions of 
crossing times 
occurs when the systems are still rather diffuse. This is due to the fact that 
the harmonic radius $R_{\rm H}$ is more sensitive to smaller separations than 
$R_{\rm S}$.

The kinematical properties of $g27c$ and $g02c$ at the present epoch are in 
good agreement with the values found in NCfA, GCF, and MK groups. In general, our simulations reproduce very well the kinematical parameters of `normal' small groups of galaxies (see Table~\ref{tab:res-tab2}).

{\bf Group $g27c$.} This group has properties at the present epoch that resemble 
those of a `normal' diffuse group: $R_{\rm H} \approx R_{\rm S} \approx 700$ kpc, 
$\sigma \approx 70$~km~s$^{-1}$, and 
$H_0 \tau_{\rm c} \approx H_0 t_{\rm c} \approx 0.5$.~These values 
are large with respect the observed ones in HCG, but in 
better agreement with the medians of NCfA, GCF, and MK groups.

{\bf Group $g02c$.} This system has a similarity with Hickson's groups; it 
shows
a $\sigma \approx 200$~km~s$^{-1}$ and  $H_0 t_{\rm c} \approx 0.09$ when is 
observed along the Y-axis (see Fig.~\ref{fig:3groups_kin}). However, its mean 
separation is quite large $R_{\rm S}\approx 400$~kpc. The average values, over 
the three lines of sight, obtained at the present epoch are: 
$R_{\rm H} \approx 10$~kpc,  $R_{\rm S}\approx 250$~kpc, $\sigma 
\approx 90$~km~s$^{-1}$,  $H_0 \tau_{\rm c} \approx 0.01$, and 
$H_0 t_{\rm c} \approx 0.2$.

{\bf Group $g22c$.}  This group has dynamical properties, at the present 
epoch, 
that resemble very closely the medians of Hickson's groups: 
$R_{\rm H} \approx 40$~kpc,  $R_{\rm S}\approx 110$~kpc, 
$\sigma \approx 120$~km~s$^{-1}$,  $H_0 \tau_{\rm c} \approx 0.1$, 
and $H_0 t_{\rm c} \approx 0.05$. However, a 
$R_{\rm S}\approx 50$~kpc, $\sigma \approx 180$~km~s$^{-1}$ 
and $H_0 t_{\rm c} \approx 0.01$ are found along the Z-axis. We should mention 
that the selection of $g22c$ was based solely on its similarity with  
HCG at the present time when seen in the XY-plane, which is confirmed 
with the more realistic models 
of spiral galaxies in section $\S 6$. In this group, the three  closest 
galaxies merge in $\sim 1$ Gyr from the present epoch, while 
the other two form 
a wide `binary' at the end of the simulation ($10$~Gyr). This result suggests 
that although some HCG have small crossing times they will not 
merge completely 
over several Gyr from the present epoch.

\begin{figure}
\epsfxsize=8.3cm
\begin{center}
\epsffile{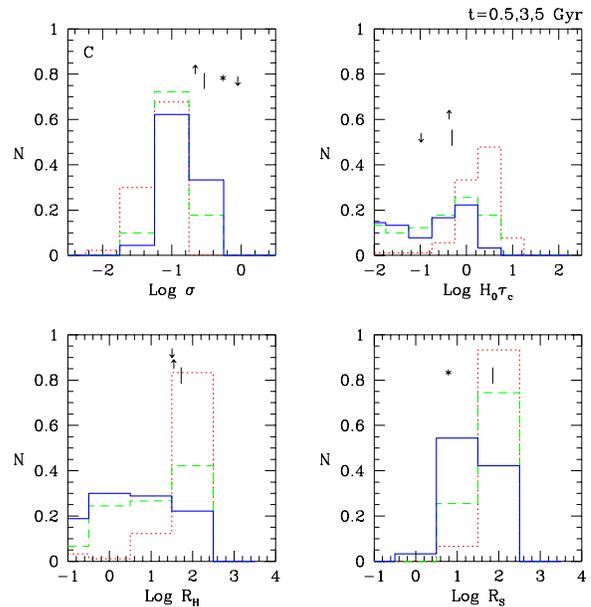}
\end{center}
\vspace{-0.7cm}
\caption{ Collapsing groups: Histograms for the velocity dispersion, 
mean harmonic radius, dimensionless crossing time ($H_0 \tau_{\rm c}$), 
and mean separation at different times. The different symbols 
denote the median 
for HCG ($\ast$), NCfA ($\mid$), and MK groups ($\uparrow$). 
Hickson Compact Groups Associations' values (Ramella et al.~1994) 
are also indicated ($\downarrow$) for completeness. Different line styles 
correspond to different times of evolution: 0.5 Gyr (dotted), 3.0 Gyr 
(short-dashed), and 5 Gyr (solid). }
\label{fig:hclpse}
\end{figure}

\begin{figure*}
\centering
\begin{center}
\epsfxsize=8cm
\epsffile{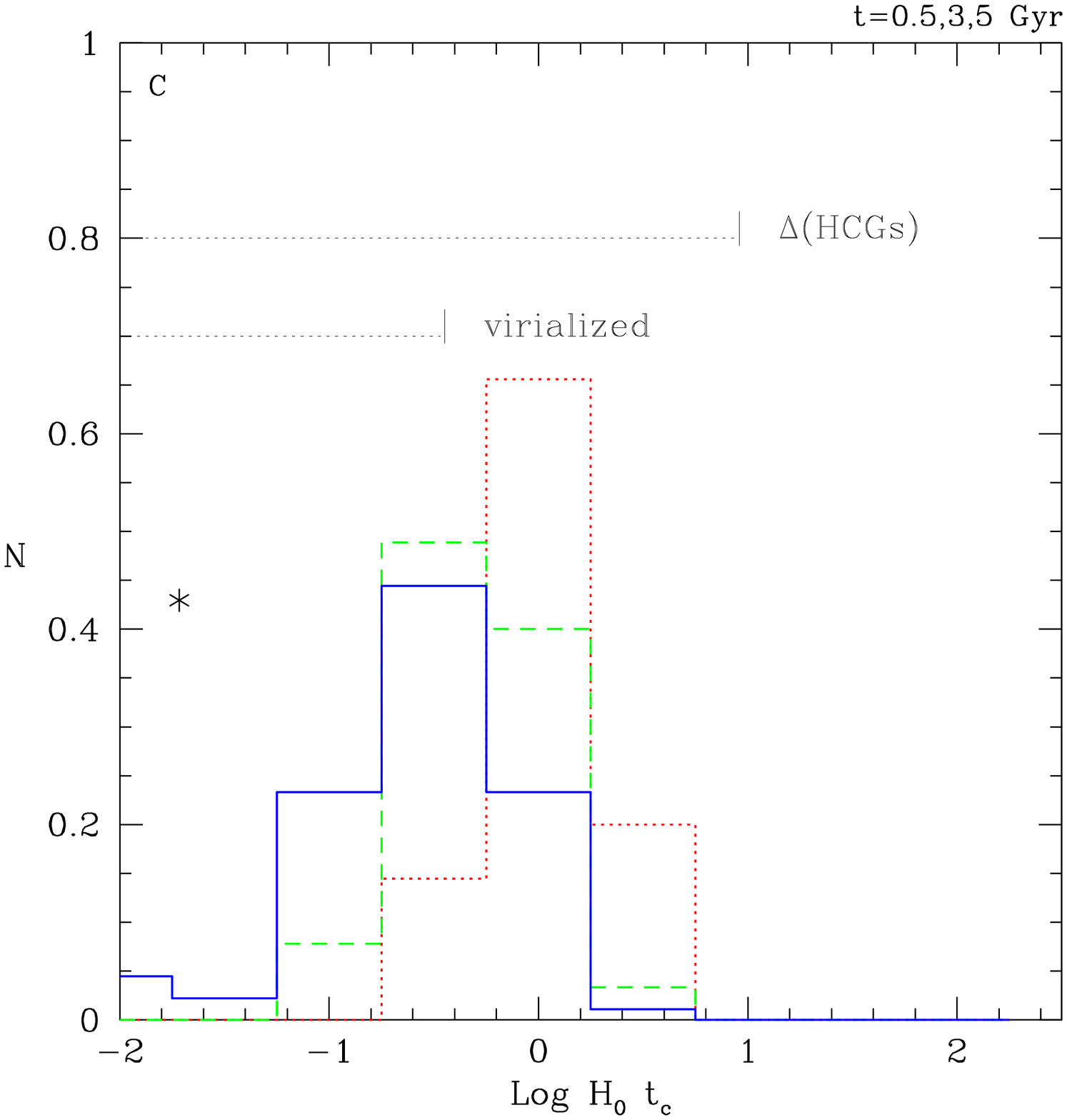}
\epsfxsize=8cm
\epsffile{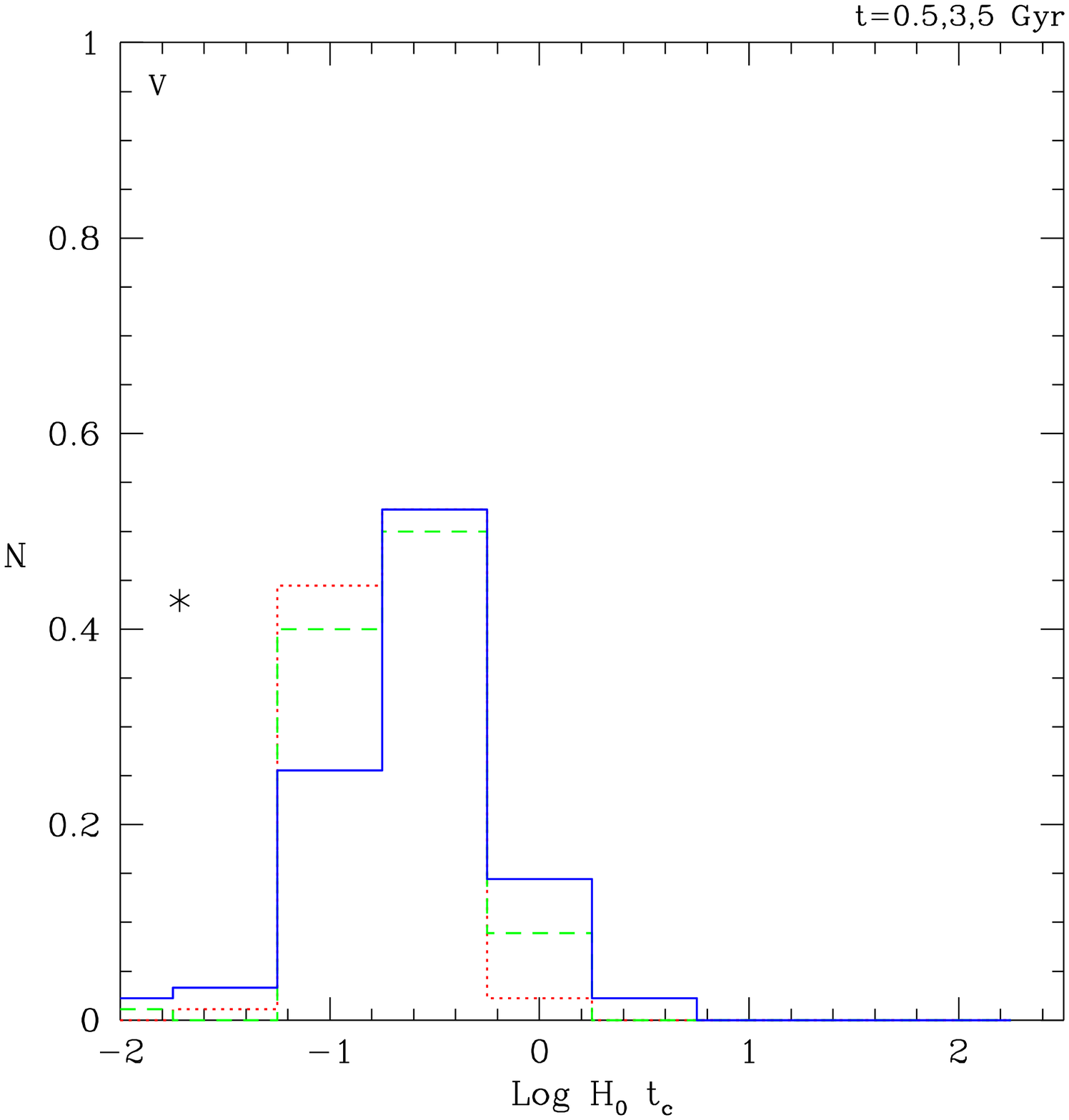}
\end{center}
\vspace{-0.3cm}
\caption{Evolution of the distribution of the crossing-time 
$t_{\rm c}=R_{\rm S}/(\sqrt{3}\sigma)$. Lines styles are the same as in 
Fig.~\ref{fig:hclpse}.  The interval of crossing-times 
for HCG and that assumed to be satisfied by  virialized groups, 
according the criterion of Gott \& Turner (1977), are indicated. 
The left panel shows results for collapsing groups, while the right one 
for virialized. Notice that both kinds of initial conditions lead to ranges 
in crossing-time within the observed values.}
\label{fig:hthickson}
\end{figure*}

Histograms for the dynamical properties of collapsing groups at times  
$t \approx 0.5,3,5$ Gyr, normalized to the initial number of 
`observed' groups, are shown in Figure~\ref{fig:hclpse}. The median values 
obtained by different authors are also indicated. The crossing time as 
defined by Hickson et al. (1992) is shown in the left-panel of 
Figure~\ref{fig:hthickson}. 

This ensemble of simulations shows a general trend to lower their size  and 
increase their velocity dispersion, until the point where `binaries' or 
`triplets' start to form. At this stage the velocity dispersion of the group 
begins to decrease due to the transfer of orbital energy to the internal 
degrees of the individual galaxies. Also, the harmonic radius 
distribution shows a strong increase in its lower end while the velocity 
distribution increases its higher end (see Fig.~\ref{fig:hclpse}). Notice that 
at the present epoch ($t\approx 5$ Gyr) there is a significant fraction of 
groups with $\sigma \gta 100$~km~s$^{-1}$.

It is important to point out that the median values observed in catalogs of 
`normal' small groups are well reproduced by our collapsing groups at the 
present epoch (e.g. $\sigma$ and $R_{\rm S}$ in Figure~\ref{fig:hclpse}). 
About $30$\% of the collapsing groups have a $\sigma$ value close to the 
median of HCG and $\approx 50$\% have a similar $R_{\rm S}$. Furthermore, 
the majority of the crossing times fall within the values found in HCG 
(Figure~\ref{fig:hthickson}) and $\approx 50$\% would be classified to be 
virialized at the present epoch accordingly with the criterion of 
Gott \& Turner (1977).

Hickson's catalog (Hickson 1982, Hickson et al. 1992) has 16 groups with 
$\sigma \lta 100$ km s$^{-1}$ and just five of them show a $\sigma \lta 5$ 
km s$^{-1}$. Four of these last groups (HCG38, HCG47, HCG49 and HCG88)
have a dimensionless crossing time of $H_0 t_{\rm c} \gta 1$ suggesting 
they are not close to virial equilibrium. However, our results indicates that 
such a assumption may not be correct. 

Since our models roughly agree with those HCG exhibiting a low 
velocity dispersion, they suggest a probable explanation 
for that fact (Rubin et al. 1991, Mamon 2000).

\begin{figure}
\epsfxsize=8.3cm
\begin{center}
\epsffile{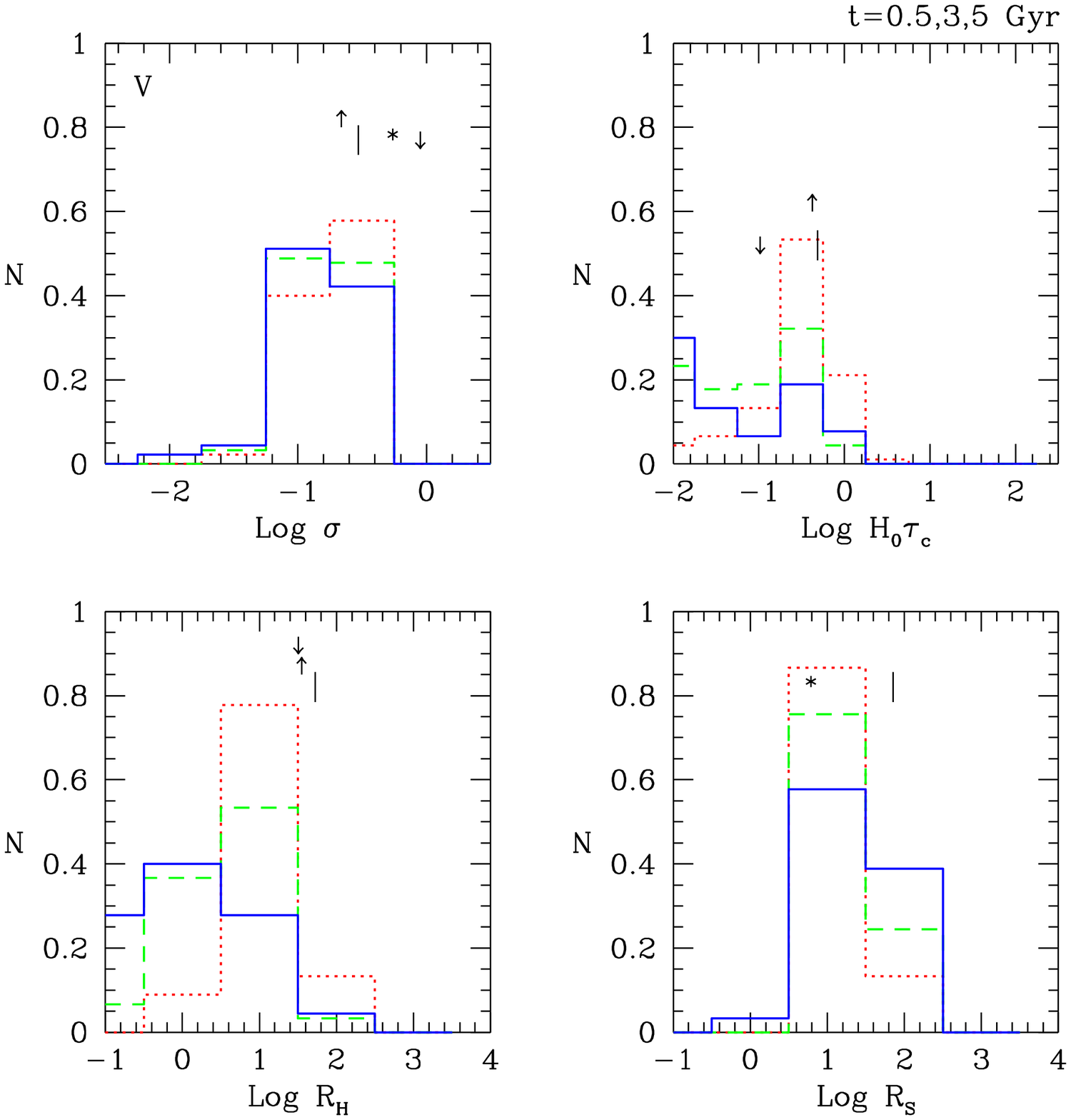}
\end{center}
\vspace{-0.7cm}
\caption{Same as in Fig.~\ref{fig:hclpse}, but for small groups with virial initial conditions.}
\label{fig:hvirial}
\end{figure}

\subsection{Virialized Groups}

Figure~\ref{fig:hvirial} contains histograms for the dynamical quantities for 
virialized initial conditions at times $t\approx 0.5,3,6$ Gyr. According 
with Table~\ref{tab:res-tab2}, these groups present a median of 
$\sigma \approx 90$~km~s$^{-1}$, $R_{\rm H} \approx 200$~kpc, 
$R_{\rm S}\approx 250$ kpc, $H_0 \tau_{\rm c} \approx 0.20$ and 
$H_0 t_{\rm c} \approx 0.13$ at their initial stages.
After $5$~Gyr of evolution the following medians are obtained:
$\sigma \approx 58$~km~s$^{-1}$, $R_{\rm H} \approx 6$~kpc, 
$R_{\rm S}\approx 274$ kpc, $H_0 \tau_{\rm c} \approx 0.01$ and 
$H_0 t_{\rm c} \approx 0.21$  which are far from the values 
found in Hickson's groups.

In general, the $\sigma$-distribution shows small changes even after $\sim 10$ 
Gyr of evolution; however, the $R_{\rm H}$-distribution presents significant 
changes which are also observed in the evolution of the 
$H_0 \tau_{\rm c}$-distribution.~The $\sigma$-distribution tends to populate the regions of lower velocity dispersion and to increase 
somewhat the mean separation. Notice that about $50$\% of groups have at
$t=0$ crossing times similar to those of normal small groups (NCfA, GCF, and MK) and  $\sim 5$\% close to the median of Hickson's groups.

The results of this and the previous sections indicate that a large fraction 
($\sim 40$\%) of our initially virialized groups have not merged during 
$10$ Gyr and, hence, the overmerging (Hickson~1997) may not be a problem. 
 However, it is important to point out that in our 
simulations a primordial common dark halo has not been included, which 
may change our conclusions.

\section{Mass Estimates}

The evolution of the virial mass (VME) and median mass (MME) 
estimators is computed. These estimators  are defined, respectively, as 
(Heisler, Tremaine \& Bahcall 1985, 
Aceves \& Perea 1999 and references therein):
$$
M_{\rm v} = \frac{3 \pi N_{\rm g}}{2 G}\frac{\sum_i V_i^2}{\sum_{i<j} 1/R_{ij}}
$$
$$
M_{\rm med} = \frac{6.5}{G} {\rm MED}_{ij} [(V_i - V_j)^2 R_{ij}]\;,
$$
where $N_{\rm g}$ is the number of galaxies, $V_i$ the velocity along the line-of-sight with respect to the centroid of velocities, $R_{ij}$ the projected separation on the sky, and the constant in $M_{\rm med}$ is determined from numerical experiments.

\begin{figure*}
\begin{center}
\epsfxsize=14cm
\epsffile{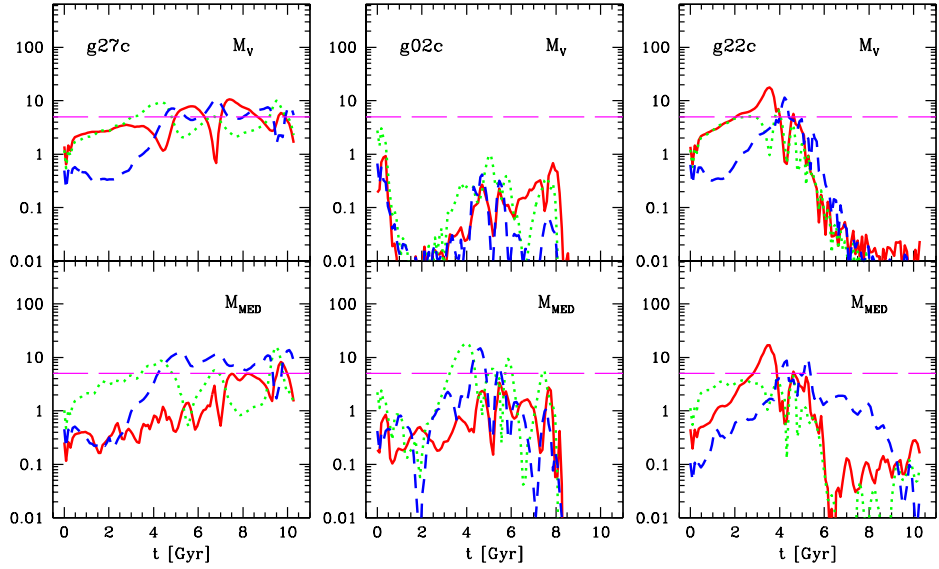}
\end{center}
\vspace{-3.7cm}
\caption{Time behavior of the two mass estimators considered in $\S 5$ for our collapsing groups $g27c$, $g02c$ and $g22c$. The upper panels show the virial mass while the bottom the median mass. The horizontal broken line corresponds to the total mass of the groups; $N$-body units are used for the mass while physical units for the time. The different types of line denote different lines-of-sight as in Figure 4.
}
\label{fig:3groups_mass}
\end{figure*}

In Figure~\ref{fig:3groups_mass} the above quantities are plotted as function 
of time for our collapsing groups $g27c$, $g02c$ and $g22c$. This figure 
reveals that both mass estimators underestimate the mass of the group 
at $t=0$ as expected (remember that these numerical groups started with 
a virial coefficient far away from equilibrium $2T/|W|=1/4$). As the group 
evolves and reaches some state of virial equilibrium it is observed that 
{\sl these mass estimators give a reliable mass for the groups as long as 
 no mergers occur between their galaxies members}. Notice that 
only bulk motions are considered in their definitions ignoring the 
self-gravitating nature of the individual galaxies. This last point is 
important since self-gravitating galaxies have the ability to absorb 
energy as the whole group evolves. In general, the VME method is more 
sensitive to the dynamical state of the system than its mass, while the MME 
is dominated by the large median separations and low velocity dispersions. 

\begin{figure*}
\centering
\epsfxsize=7cm
\epsffile{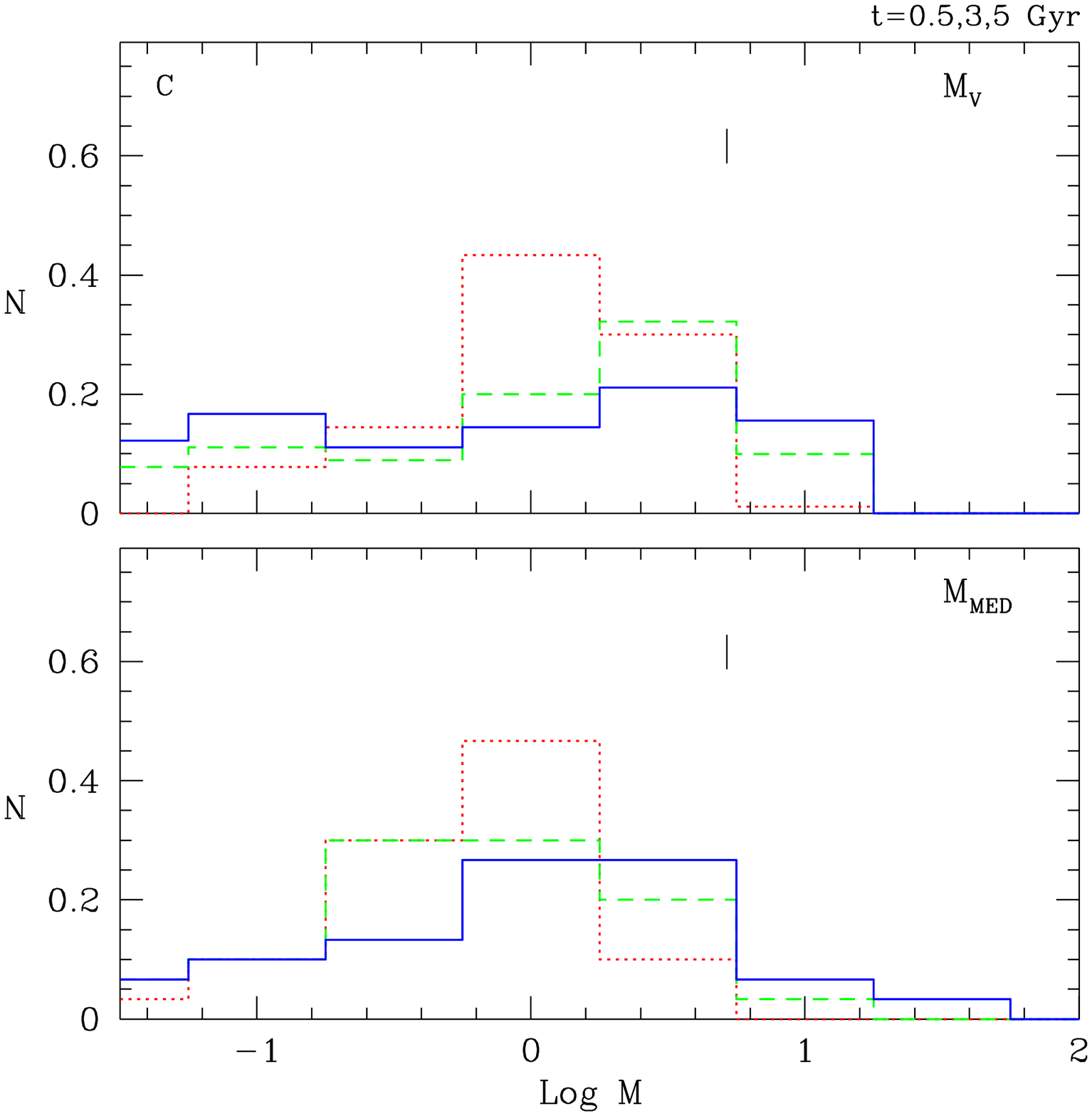}
\epsfxsize=7cm
\epsffile{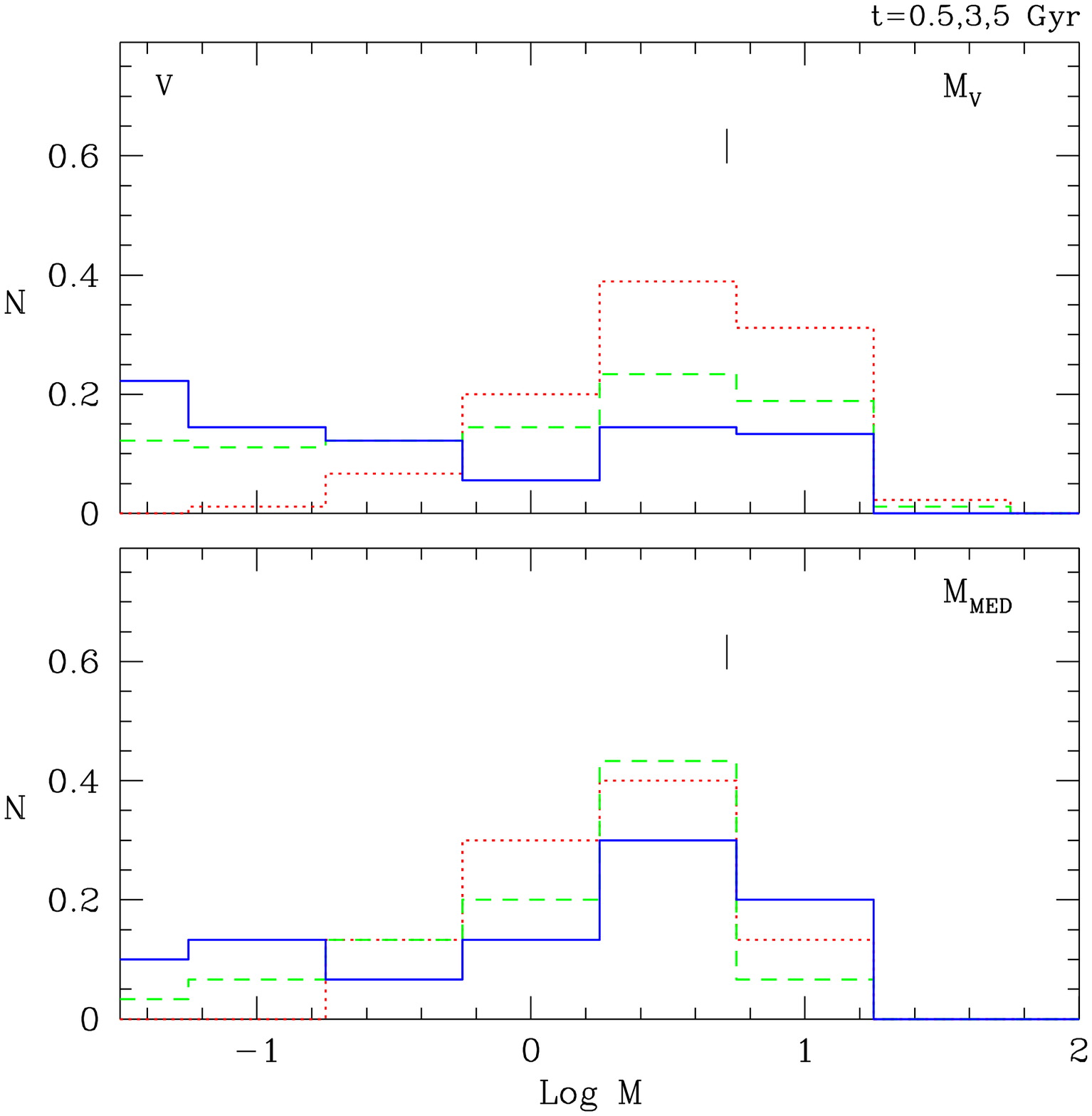}
\vspace{-0.3cm}
\caption{Histograms of the behavior of the VME ({\it upper panels}) and 
MME ({\it lower panels}), both for cold ({\it left}) and virial ({\it right}) 
small groups. The bar ($\mid$) indicates the true mass of the groups in 
$N$-body units. Dotted lines show the distribution at $t=0.5$ Gyr, 
short-dashed lines at 3 Gyr and solid at 5 Gyr. }
\label{fig:hmasas-hist}
\end{figure*}

The results from all the simulations carried out are summarized in 
the histograms of Figure~\ref{fig:hmasas-hist}. The true mass of the groups 
is indicated by the vertical line. For collapsing groups (left panels) the 
VME estimator (left-upper panel) gives a median virial mass of 
$M_{\rm V}=1.28$ at $t=15$ which is consistent with 
the expected value of $M_{\rm V}=1.25$,\footnote{\footnotesize
 According to Bahcall \& Tremaine 
(1981) and Geller (1984) a lower fractional limit on the virial mass is 
 $\delta_{\rm v} \gta \sqrt{2 \ln N}/(\pi^2 \sqrt{N})$, leading to $M_{\rm v}
\approx 1.28 \pm 0.12$ for $N=90$ `observed'\/ groups.}
while the MME (left-lower panel) underestimates the 
mass by about an order of magnitude. However, once the groups are allowed 
to evolve to the present epoch ($\sim 5$ Gyr) the mass is underestimated 
by an order of magnitude for the VME and by a factor of about $5$ for the 
MME method. For the case of virialized groups we found that at $t=0$ 
both mass estimators give a similar median mass of about 
$M_{\rm V}\approx 3.5$. Once  the separation between galaxies 
decreases, the MME provides a better estimate of the mass. 
From these results, the MME is considered more appropriate to estimate the mass of systems that 
resemble HCG.

Hickson's compact groups have a mass-to-light ratio of $\Upsilon \approx 50h \Upsilon_{\odot}$. If HCG are {\it physically}  well defined groups, their 
content of matter is probably $\gta 5$ times more than is inferred now. 
This would 
imply that a closer estimate to the true mass-to-light ratio of HCG would 
be $\Upsilon \sim 250 h \Upsilon_\odot $. It is interesting to note that 
this value falls within the values found in clusters 
$200-300 h \Upsilon_{\odot}$ \cite{neta99} suggesting that both types of 
systems would have approximately the same fraction of baryonic to dark matter 
(see White~et~al.~1993 for a discussion of the cosmological implications).

\section{$R_{\rm S}$--$\sigma$ Diagram}\label{sec:evolution}

\begin{figure*}
\centering
\epsfxsize=55.5mm
\epsffile{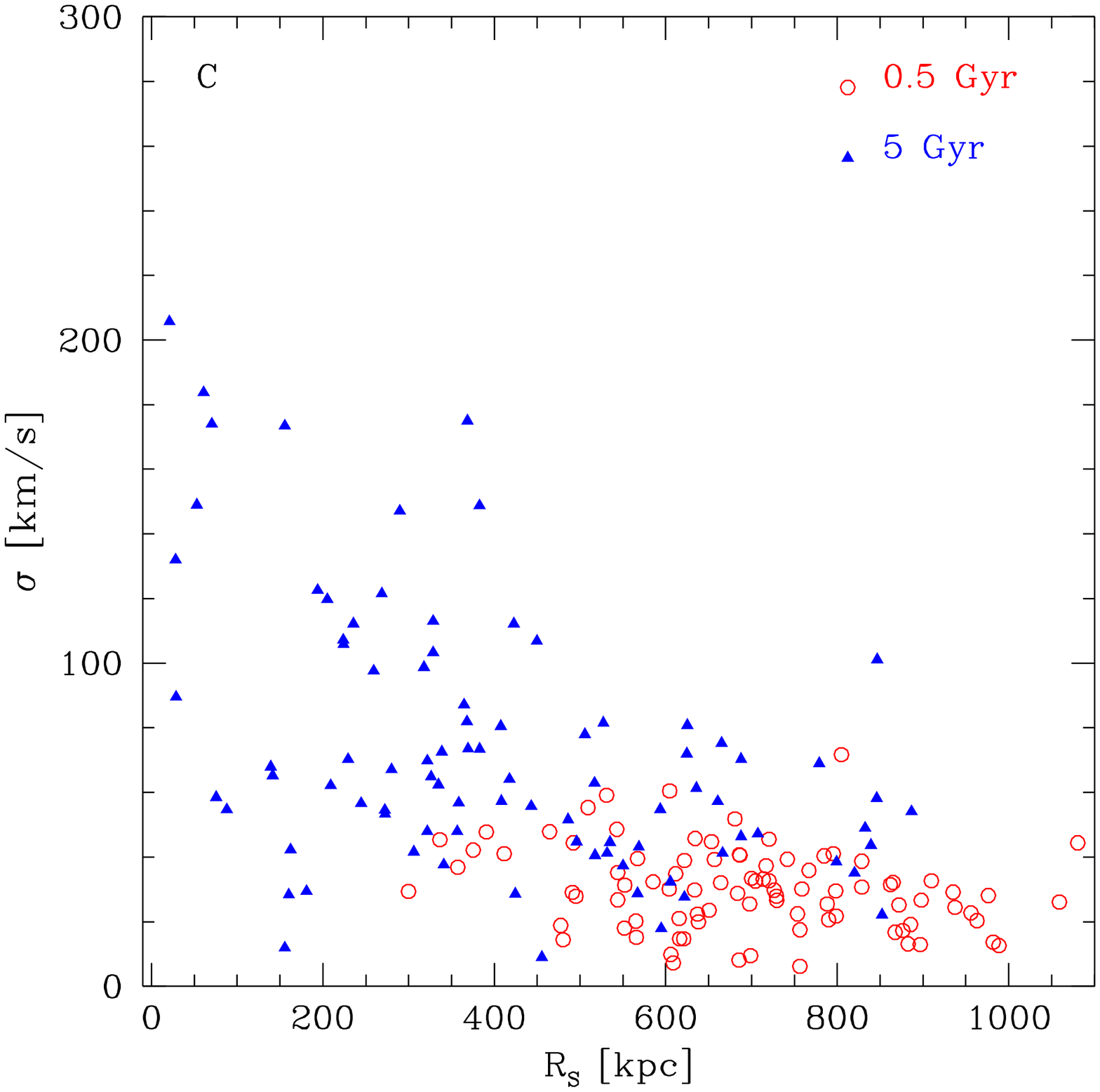}
\epsfxsize=55.5mm
\epsffile{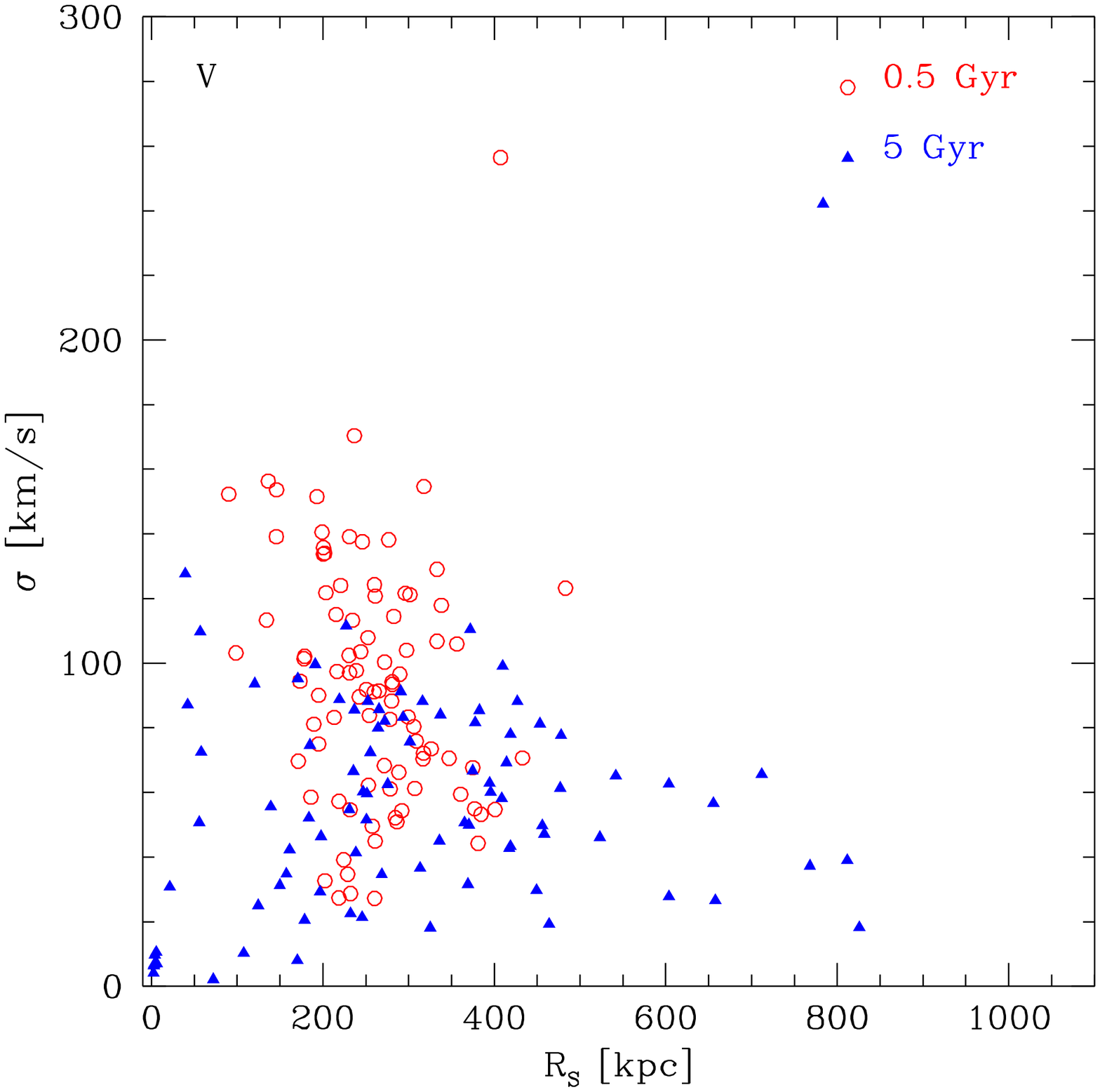}
\epsfxsize=55.5mm
\epsffile{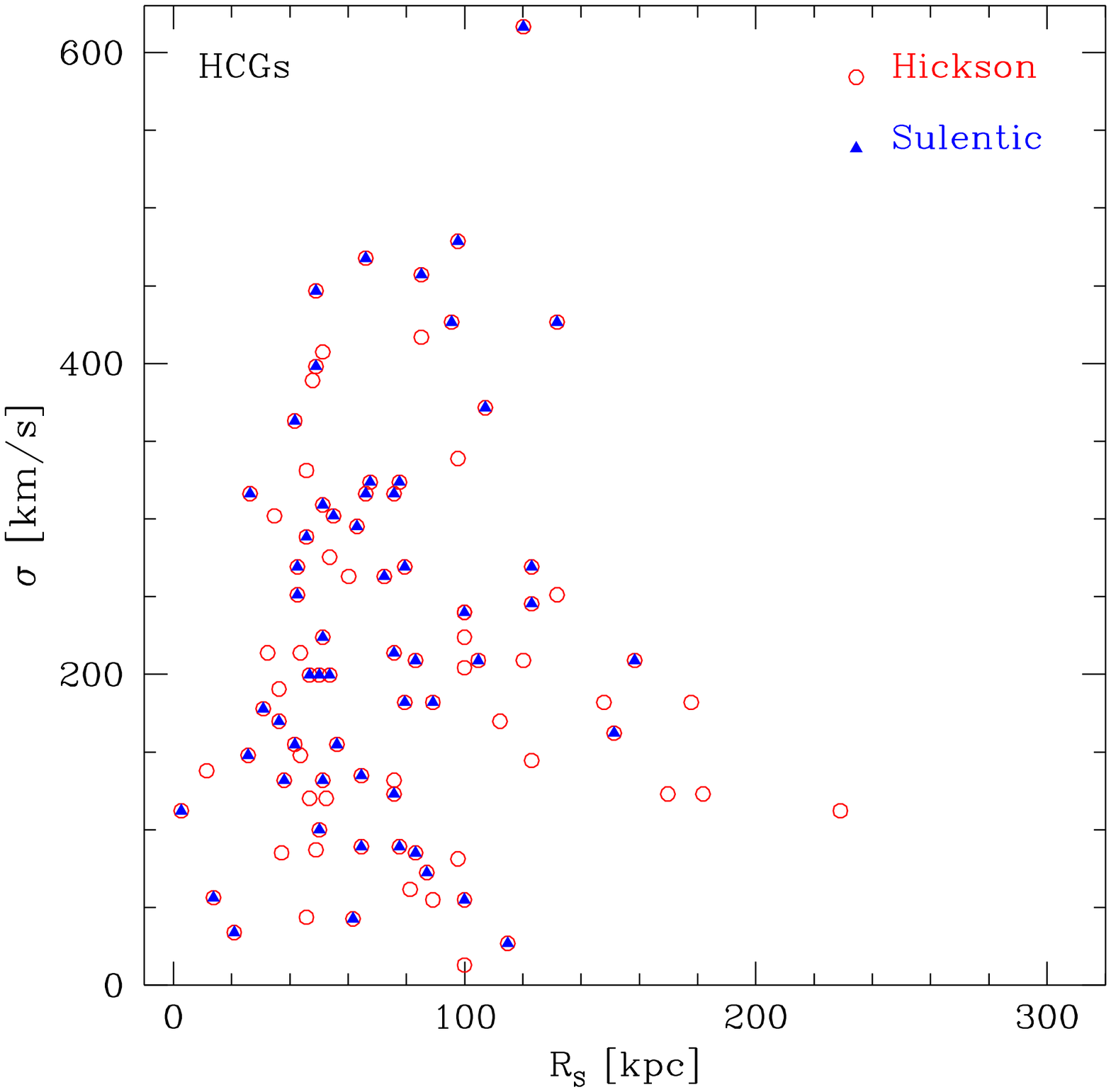}
\vspace{-0.8cm}
\caption{$R_{\rm S}$-$\sigma$ diagram for collapsing ({\it left}), virialized 
({\it center}), and Hickson's groups ({\it right}). Hickson's compact groups 
appear to resemble systems that have suffered a gravitational collapse out 
of a more diffuse system. }
\label{fig:phase-planes}
\end{figure*}

A restricted `phase-space' consisting of a $R_{\rm S}$--$\sigma$ 
diagram is constructed in order to try to discern some evolutionary track in our simulated 
groups. In Figure~\ref{fig:phase-planes} the collapsing (left-panel) 
and virialized (center-panel) groups are shown at times $t=0.5$ and $t=5$ Gyr together 
with the corresponding diagram for HCG where the 
ones considered `bona fide' by Sulentic (1997) are indicated.
  
This diagram clearly shows that initially cold groups reduce their mean 
separation and increase their velocity dispersion as they evolve, while 
for virialized groups their velocity dispersions decrease somewhat and their 
mean separation tends to increase. 
	Furthermore,  about $8$ collapsing groups have $R_{\rm S} \lta 100$~kpc and $\sigma \in (60,200)$~km~s$^{-1}$ at the present epoch ($\sim 5$ Gyr) which roughly 
correspond to the values shown by `normal' small groups, and are within the range of values found 
for HCG.  Also, $2$ groups have $R_{\rm S} \lta 50$~kpc 
and $\sigma \approx 200$~km~s$^{-1}$ which are very close to the median 
values of HCG.

This $R_{\rm S}$-$\sigma$ diagram suggests a resemblance of HCG with our the 
systems that once had a diffuse configuration and are at the verge of 
collapse at the present time, while some of our virialized groups show 
properties more similar to the `normal' small groups.      
These and previous results suggest that HCG are the more advanced systems in 
a process of gravitational collapse among small groups. Thus, compact groups 
may be rather young configurations (Hickson~et~al.~1984, Barnes 1989, White 1990, Rubin~et~al.~1991).

\section{A more realistic simulation}

\begin{table*}[t]
\caption[]{Galaxy parameters}
\label{tab:spiral}
\begin{center}
{\small
\vspace{-0.3cm}
\begin{tabular}{cccccccccccccc}
\hline\noalign{\smallskip}
Model & $M_D$ & $R_D$ & $R_t$ & $z_D$ & $\delta R_{out}$ & $\Psi_c$ & $\sigma_B$ & $\rho_B$ & $\Psi_0$ & $\sigma_0$ & $q$ & $C$ & $R_a$ \\
\noalign{\smallskip}
\hline
\noalign{\smallskip}
 MW-B & 0.87 & 1.0 & 4.0 & 0.15 & 0.4 & -2.9 & 0.71 & 14.5 & -5.2 & 0.96 & 1.0 & 0.1 & 0.8 \\ 
\noalign{\smallskip}
\hline
\end{tabular}
}  
\end{center}
{\small
{\scriptsize${M_D}$}, {\scriptsize $R_D$}, {\scriptsize${R_t}$},  
{\scriptsize$z_D$} and {\scriptsize $\delta R_{out}$} correspond to the mass, 
radial scale-length, cut-off radius, vertical scale-length of the disk, and disk truncation width respectively. 
 {\scriptsize$\Psi_c$}, {\scriptsize$\sigma_B$} and 
{\scriptsize$\rho_B$} refer to the cutoff potential, velocity dispersion, and central density of the bulge, respectively.
	 Finally, {\scriptsize$\Psi_0$}, 
{\scriptsize$\sigma_0$}, {\scriptsize$q$}, {\scriptsize$C$} and 
{\scriptsize$R_a$} 
indicate the central potential, velocity dispersion, potential flattening, concentration, and characteristic radius of the halo, respectively.
}
\end{table*}

In the simulations reported above no clear distinction between luminous and 
dark matter was  made, and hence no precise statement could be established whether, 
for example, the group $g22c$ could actually be considered a compact group 
according to Hickson's criteria. 

\begin{figure}
\centering
\epsfxsize=80mm
\epsffile{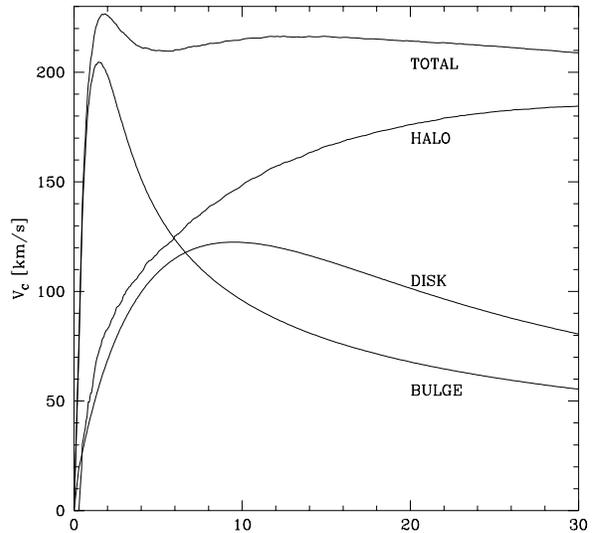}
\vspace{-1cm}
\caption{Rotation curve for the spiral galaxy model.} 
\label{fig:rotcurve}
\end{figure}

To address this particular issue a more realistic galaxy model was set up by   
replacing the plummer models by spiral galaxies using model B of Kuijken \& 
Dubinski (1995); see $\S$2.1 for some numerical values of this model (we refer the reader to Kuijken \& Dubinski's paper for a detailed description 
of the method to build up the disk galaxy model). It should be mention that 
a detailed study of groups involving disk galaxies is out the scope of the 
present work. 
Table~\ref{tab:spiral} 
summarizes the parameters that define our galaxy model and 
Figure~\ref{fig:rotcurve} shows its corresponding rotation curve. 

For the galaxy centers and bulk motions inside the group the 
values of our previous collapsing model $g22c$ are used. For illustrative purposes, arbitrary orientations for the galaxy angular momentum vector are chosen. 
Each numerical galaxy consists of 65536 particles, 16384 for the disk, 4096 
for the bulge and 45056 for the halo; i.e., a total of 327680 
were used in this simulation.

\begin{figure*}
\centering
\epsfxsize=80mm
\epsfxsize=80mm
\caption{({\it left}) XY-projection at time 4.8 Gyr of our more realistic 
 model. Luminous particles are shown in the {\it left-panel} and  
the dark component is indicated in the {\it right-panel}. Notice that a 
common dark envelope for the three galaxies close to merging is already 
present,
while the other two in the upper-left will not merge for 
$\gta 5$ Gyr from the present epoch. Compare with our previous model 
$g22c$ of Figure 1.} 
\label{fig:group}
\end{figure*}

To evolve the system a {\sc parallel treecode} developed by Dubinski (1996) with a tolerance parameter of $0.8$ was used. Forces were computed with the 
quadrupole terms included and with a fixed time-step corresponding to $1$ Myr. 
With these parameters the energy conservation was better than 0.2\%. 
The simulation was ran in a Beowulf class cluster consisting of 
32 Pentium processors of $450$ MHz (Vel\'azquez \& Aguilar 2002). 
It took a wall time of 16 seconds per time-step.

In Figure~\ref{fig:group} ({\it left}) the XY-projection of the 
luminous component of the group at time $4.8$ Gyr, the same as figure~1 of 
model $g22c$, is shown. Although differences in the evolutionary 
trend of our more realistic model are appreciated 
with respect to  model $g22c$, it is 
rather easy to conclude that it satisfies Hickson's criteria for  
compact groups. Hickson's criterion on the difference of magnitudes among 
galaxies is satisfied immediately since our galaxies are identical. 
Therefore, a group is obtained that is compact at the present epoch and 
will not merge completely for another $\sim 5$ Gyr.

In Figure~\ref{fig:group} ({\it right}) only the dark particles of 
the haloes are shown. Notice that a common dark halo has already formed 
for three of the galaxies that will merge in the next few Gyr. This result is 
consistent with the suggestion by Rubin et al. (1991) that compact groups 
reside in a common dark halo at the present epoch.

\section{General Discussion}

	It has been found that $\approx 10$\% of initially diffuse groups reach the  present epoch with dynamical properties consistent with the median of Hickson's compact groups. 
	Hickson's groups appear as those systems that have manage to evolve more rapidly toward a compact configuration, due to particular initial conditions in the density perturbation that led to the formation of small groups.  
	The formation of a compact groups is, according to our results, consistent with a hierarchical clustering scenario for the formation of structures in the universe. Hence supporting the hypothesis advanced by Barnes (1989) and White (1990), and the one suggested on observational grounds (Hickson~et~al.~1984, Rubin~et~al.~1991).

	The present model leads to $\sim 50$\% of originally diffuse groups arriving at the present epoch without any merging.
	Also, it is found  that $\sim 10$\% (5 simulations) of these groups will not have any merger within the next $\sim 5$ Gyr. In this sense, our model provides an alternative explanation to the properties of compact groups and their existence today, and complements several previous studies (e.g., Athanassoula~et~al.~1997).

	In general, initial conditions from `maximum expansion' reproduce very well the properties of `normal' small groups. 
	Our simple model avoids the requirement for an initial overdensity and/or significant secondary-infall (Diaferio~et~al.~1994, Governato~et~al.~1996), and the need for especial initial conditions to avoid the complete merging of a group (e.g., Athanassoula~et~al.~1997).

	The present model does not show the merging instability, but many questions remain.
	One of the most obvious caveats of our simulations is not taking into account a primordial dark halo for virialized initial conditions. However, the merging activity found  may still  be considered a lower limit if a concentrated dark halo were to be included. Surely more realistic simulations of small groups, considering a spectrum of masses and a clear distinction between luminous and dark matter, as well as initial conditions taken from a cosmological simulation, are necessary to address these issues. We plan to study some of these matters in the future.

\section{Conclusions}

	The main results of this work are as follows:

\begin{enumerate}

\item  Groups starting from `maximum expansion' with a diffuse configuration 
have, at the present epoch, dynamical properties very similar to the medians obtained 
from several catalogs of small groups. In particular, about $10$\% of them 
show values close to the observed medians of Hickson's groups.

\item The suggestion that Hickson's compacts groups originated from diffuse 
systems, and are relatively young systems, finds quantitative support in our numerical experiments.

\item It is found that overmerging is not an important 
problem in our 
simulations, either  with virialized or cold initial conditions. For the case of virialized 
groups about $40$\% survive merging during $\approx 10$ Gyr. This also 
indicates that 
the existence of a large, massive common dark halo does not appear to be a 
strong requirement to explain the present day properties of small groups. 

\item The median mass estimator (MME) appears to be a better 
estimator of mass than the virial mass estimator (VME) for groups at 
the present epoch. It is estimated that the mass-to-light ratio 
of HCG is probably $\Upsilon \sim 250 \Upsilon_\odot$ suggesting that clusters 
and compact groups have about the same fraction of baryonic to dark matter.

\end{enumerate} 

\section*{Acknowledgments}  
H.A. thanks CONACyT (M\'exico, Proyecto:37506-E) and the Spanish Ministry of Foreign Affairs 
 for financial support. H.A. thanks greatly Gary Mamon, 
Jack Sulentic, Lourdes Verdes-Montenegro, and Jaime Perea  for important 
input on this 
work, and P. Hickson and P. Fouqu\'e for sending their groups' catalogs. 
H.V. was supported by a grant from CONACyT (Proyecto:27678-E). We thank W. J. Schuster for a careful reading of the final manuscript.
Finally, we thank the anonymous referee for suggestions to improve and clarify this work.



\end{document}